\documentclass[doublecol]{epl2b}
\usepackage{graphicx,amsmath,amssymb,amsfonts,latexsym,color,dcolumn,bm,epsfig,subfigure}
\usepackage[latin1]{inputenc}

\def\bbm[#1]{\mbox{\boldmath $#1$}}
\newcommand{\ket}[1]{\displaystyle{|#1\rangle}}
\newcommand{\bra}[1]{\displaystyle{\langle #1|}}

\newcommand{\Rea}{\text{Re}}
\newcommand{\Ima}{\text{Im}}
\newcommand{\TE}{\text{TE}}
\newcommand{\TM}{\text{TM}}

\title{Dynamics of an elementary quantum system in environments out of thermal equilibrium}
\shorttitle{Dynamics of an elementary quantum system in environments out of thermal equilibrium} 

\author{B. Bellomo\inst{1,2}\thanks{E-mail: \email{bruno.bellomo@univ-montp2.fr}}, R. Messina\inst{3} \and M. Antezza\inst{1,2}}
\shortauthor{B. Bellomo \etal}

\institute{
  \inst{1} Universit\'{e} Montpellier 2, Laboratoire Charles Coulomb UMR 5221 - F-34095, Montpellier, France, EU\\
  \inst{2} CNRS, Laboratoire Charles Coulomb UMR 5221 - F-34095, Montpellier, France, EU\\
  \inst{3} Laboratoire Charles Fabry, Institut d'Optique, CNRS, Universit\'{e} Paris-Sud, Campus Polytechnique, RD 128, 91127 Palaiseau cedex, France, EU}


\abstract{We study the internal dynamics of an elementary quantum system placed close to a body held at a temperature different from that of the surrounding radiation. We derive general expressions for lifetime and density matrix valid for bodies of arbitrary geometry and dielectric permittivity. Out of equilibrium, the thermalization process and steady states become both qualitatively and quantitatively significantly different from the case of radiation at thermal equilibrium. For the case of a three-level atom close to a slab of finite thickness, we predict the occurrence of population inversion and an efficient cooling mechanism for the quantum system, whose effective internal temperature can be driven to values much lower than both involved temperatures. Our results show that non-equilibrium configurations provide new promising ways to control the state of an atomic system.}

\begin{document}

\maketitle

\section{Introduction}

Thermalization mechanisms in quantum systems driven by changes of external parameters offer a great variety of relaxation phenomena, typically studied for many-body systems \cite{RigolNature,PolkovnikovRMP,EisertPRL,AltlandPRL,TrotzkyNP}. What happens to the internal dynamics of an elementary one-body quantum system in presence of an environment driven out of thermal equilibrium? This configuration characterizes several systems in biology \cite{CaiPRE10} and physics \cite{AntezzaPRL05,AntezzaJPA06,ObrechtPRL07,BrunnerPRE12,EmmertEPJD,Camalet2011} and, despite its simplicity, it may offer a great richness. Systems out of thermal equilibrium have been recently subject of intense investigations concerning heat transfer \cite{RousseauNP,ShenNL,Ben-AbdallahPRB2010,OttensPRL,MesAntEPL11,MesAntPRA11,KardarPRL,RodriguezPRL,Messina2012} and Casimir-Lifshitz interaction \cite{AntezzaPRL05,ObrechtPRL07,AntezzaPRLA06,AntezzaPRA08,BuhmannPRL08,BehuninPRA2011,MesAntEPL11,MesAntPRA11,SherkunovPRA09}.
There,  the  thermalization dynamics is not considered, being the full system assumed fixed in a given configuration.

The key quantity for the internal atomic dynamics is the lifetime. Approaching an atomic system close to a body modifies the local electromagnetic (EM) field surrounding it, resulting in lifetimes depending on their relative distance, on geometrical and optical properties of the body, as well as on the temperature of the system. Typically lifetime has been studied for configurations at thermal equilibrium, when the radiation impinging on the body is at thermal equilibrium with it \cite{Novotny06,BiehsPRA11,BenAbdallaharXiv11}. Such studies have particular relevance in atom-chip experiments \cite{ColombeNature07,FortagRevModPhys07} and in all experiments where the near field of a body is probed \cite{AntezzaPRL05,KittelPRL05,DeWildeNature06,ObrechtPRL07,KawataNaturePhotonics09}.

\begin{figure}[h!]
\includegraphics[width=0.46\textwidth]{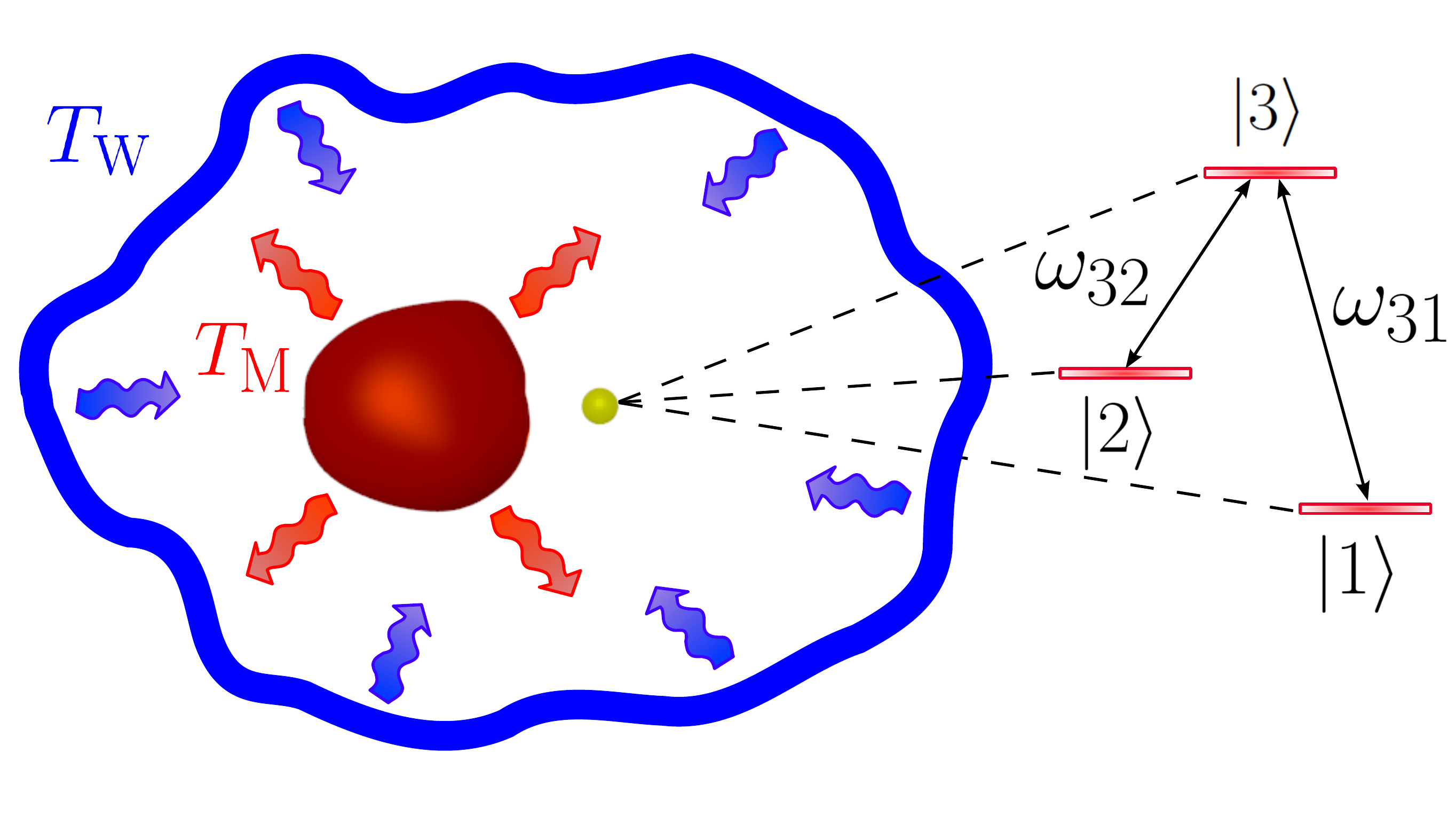}
\caption{\label{fig:1}\footnotesize The temperature of the body, $T_\mathrm{M}$, and the one of the surrounding walls located at large distance, $T_\mathrm{W}$, may be different, and are kept fixed in time realizing a stationary configuration out of thermal equilibrium. For the three-level atom, having position $\mathbf{R}=(\mathbf{r},z)$ and states of increasing energy $\ket{1}, \ket{2}$ and $ \ket{3}$, we consider a $\Lambda$ configuration with only two allowed dipole transitions between $\ket{1}$ and $\ket{3}$ and between $\ket{2}$ and $\ket{3}$, respectively separated by energies $\hbar\omega_{31}$ and $\hbar\omega_{32}$.}
\end{figure}

Here we study an elementary quantum system offering a rich non-equilibrium dynamics: a three-level atomic system placed close to a body of arbitrary geometry and dielectric permittivity, held at temperature $T_\mathrm{M}$, and embedded in an additional radiation coming from the walls surrounding the entire system and held at temperature $T_\mathrm{W}$ (see Fig. \ref{fig:1}). When the walls are irregular and far enough from the atom-body system, their contribution can be described as a black-body radiation in the region of the system \cite{AntezzaPRL05}. What happens to lifetimes, coherences, and steady states? How are these quantities modified with respect to the thermal equilibrium configuration? What is the role of geometric and dielectric properties of the body? In this letter, we answer such questions, which have both fundamental nature and experimental relevance for systems which are naturally out of thermal equilibrium, such as recent studies of cold
atoms close to superconducting surfaces \cite{GuerlinNature07,SayrinNature11}, or involving the tip of an AFM close to cold samples \cite{HenkelJOptA02,JoulainarXiv12}. Our analysis, concerning  both real and artificial atomic systems, can be readily extended to more complicated level schemes.

\section{Atomic dynamics}

The atom\,-\,field interaction is described using the multipolar-coupling Hamiltonian $H_I=-\mathbf{D}\cdot\mathbf{E}(\mathbf{R})$, where $\mathbf{D}$ is the atomic electric dipole operator (supposed to be purely non-diagonal and with non-zero matrix elements $\mathbf{d}_{13}=\bra{1}\mathbf{D}\ket{3}$ and $\mathbf{d}_{23}=\bra{2}\mathbf{D}\ket{3}$) and $\mathbf{E}(\mathbf{R})$ is the total electromagnetic field at the atomic position. The state of the atom is represented by its density matrix $\rho(t)$, whose exact time evolution is governed, in the interaction picture, by the trace on the degrees of freedom of the field of the von Neumann equation for the total density matrix $\dot{\rho}_{\textrm{tot}}(t)=-\frac{i}{\hbar}[H_I(t),\rho_{\textrm{tot}}(t)]$. We treat here the atomic dynamics in the limit when Born, Markovian and rotating-wave approximations hold \cite{BookBreuer}, deriving a master equation in terms of transition rates associated to the allowed transitions. The validity of these approximations
is guaranteed by the weak coupling between the atom and the field due to the small value of the electric-dipole matrix elements.
The master equation reads
\begin{equation}\label{master equation N}\begin{split}&\frac{d}{dt}\rho(t)=-i\Bigl[\sum_n{\omega_n}\ket{n}\bra{n}+\sum_{m,n}S(-\omega_{nm})\ket{m}\bra{m}\\
&+\sum_{m,n} S(\omega_{nm})\ket{n}\bra{n},\rho(t)\Bigr]\\
&+\sum_{m,n}\Gamma(-\omega_{nm})\Big(\rho_{mm}\ket{n}\bra{n}-\frac{1}{2}\{\ket{m}\bra{m},\rho(t)\}\Big)\\
&+\sum_{m,n}\Gamma(\omega_{nm})\Big(\rho_{nn}\ket{m}\bra{m}-\frac{1}{2}\{\ket{n}\bra{n},\rho(t)\}\Big),\end{split}\end{equation}
where $
\Gamma(-\omega_{nm})=\sum_{i,j}\gamma_{i j}(-\omega_{nm})[\textbf{d}_{mn}]_{i}[\textbf{d}_{mn}]^*_{j}$, $
\Gamma(\omega_{nm})=\sum_{i,j}\gamma_{i j}(\omega_{nm})[\textbf{d}_{mn}]^*_{i} [\textbf{d}_{mn}]_{j}$, $n=\{1,2,3\}$, $(nm)\in\{(31),(32)\}$ and $\omega_{31}\neq\omega_{32}$. $S(\omega_{nm})$ and $S(-\omega_{nm})$ are frequency shifts not playing any role in the population dynamics. In the previous equation, $\gamma_{ij}(\omega)$ is defined by
\begin{equation}\label{gammaijdef}\gamma_{i j}(\omega)=\frac{1}{\hbar^2}\int_{-\infty}^\infty ds\,e^{i\omega s}\langle E_i(\mathbf{R},s)E_j(\mathbf{R},0)\rangle ,\end{equation}
where we used homogeneity in time for the field correlation functions. The steady solution of this master equation gives coherences equal to zero and populations
\begin{equation}\label{steady state}\begin{split}&\begin{pmatrix}\rho_{11}(\infty)\\\rho_{22}(\infty)\\\rho_{33}(\infty)\end{pmatrix}=
\frac{1}{Z}\begin{pmatrix}n_{\mathrm{eff}}^{(32)}\bigl(1+n_{\mathrm{eff}}^{(31)}\bigr)\\\vspace{-.3cm}\\
n_{\mathrm{eff}}^{(31)}\bigl(1+n_{\mathrm{eff}}^{(32)}\bigr)\\\vspace{-.3cm}\\n_{\mathrm{eff}}^{(31)}n_{\mathrm{eff}}^{(32)}\end{pmatrix},\\
&\hspace{.5cm}Z=3n_{\mathrm{eff}}^{(31)}n_{\mathrm{eff}}^{(32)}+n_{\mathrm{eff}}^{(31)}+n_{\mathrm{eff}}^{(32)}\end{split}\end{equation}
where, for $(nm)\in\{(32),(31)\}$,
\begin{equation}\label{n effective ij}
n_{\mathrm{eff}}^{(nm)}= \frac{n(\omega_{nm},T_\mathrm{W})\alpha_\mathrm{W}(\omega_{nm})
+ n(\omega_{nm},T_\mathrm{M}) \alpha_\mathrm{M}(\omega_{nm})
}{\alpha_\mathrm{W}(\omega_{nm})+\alpha_\mathrm{M}(\omega_{nm})},
\end{equation}
with
$n(\omega,T)=\bigl(\exp[\hbar \omega/k_BT]-1\bigr)^{-1}$.
The coefficients $n_{\mathrm{eff}}^{(nm)}$ associated to the two allowed transitions satisfy $n(\omega_{nm},T_\mathrm{min})\le n_{\mathrm{eff}}^{(nm)} \le n(\omega_{nm},T_\mathrm{max})$, where $T_\mathrm{min}=\mathrm{min}\{T_\mathrm{W},T_\mathrm{M}\} $ and $T_\mathrm{max}=\mathrm{max}\{T_\mathrm{W},T_\mathrm{M}\} $.
$\alpha_\mathrm{W}(\omega_{nm})$ and $\alpha_\mathrm{M}(\omega_{nm})$
are two temperature-independent functions strictly connected to the EM field correlators
and carrying all the dependence on geometrical and material properties of the body. The correlation functions of the total field appearing in eq. \eqref{gammaijdef} can be given in terms of the correlators of the fields emitted by each source by expressing the total field in terms of the reflection and transmission operators $\mathcal{R}$ and $\mathcal{T}$  associated to the side of body on which the atom is located.  Each source can be  treated independently as if it was at thermal equilibrium at its own temperature
and the correlators of the source fields can be then characterized  by using the fluctuation-dissipation theorem \cite{MesAntPRA11}. The expressions of $\alpha_\mathrm{W}(\omega_{nm})$ and $\alpha_\mathrm{M}(\omega_{nm})$ can be thus obtained, for an arbitrary frequency $\omega=\omega_{nm}$, as a function of  $\mathcal{R}$ and $\mathcal{T}$ as
\begin{equation}\label{alphas}\begin{split}&\alpha_\text{W}(\omega)=\frac{3\pi c}{2\omega}\!\sum_{p,p'}\sum_{i,j}\frac{[\mathbf{d}_{mn}]_i^*[\mathbf{d}_{mn}]_j}{|\mathbf{d}_{mn}|^2}\int\frac{d^2\mathbf{k}}{(2\pi)^2}\int\frac{d^2\mathbf{k}'}{(2\pi)^2}\\
&\,e^{i(\mathbf{k}-\mathbf{k'})\cdot\mathbf{r}}\bra{p,\mathbf{k}}\Bigl[e^{-i(k_z-k_z^{'*})z}[\hat{\bbm[\epsilon]}_p^-(\mathbf{k},\omega)]_i[\hat{\bbm[\epsilon]}_{p'}^-(\mathbf{k}',\omega)]_j^*\mathcal{P}_{-1}^{\text{(pw)}}\\
&\,+e^{i(k_z+k_z^{'*})z}[\hat{\bbm[\epsilon]}_p^+(\mathbf{k},\omega)]_i[\hat{\bbm[\epsilon]}_{p'}^{-}(\mathbf{k}',\omega)]_j^*\mathcal{R}\mathcal{P}_{-1}^{\text{(pw)}}\\
&\,+e^{-i(k_z+k_z^{'*})z}[\hat{\bbm[\epsilon]}_p^-(\mathbf{k},\omega)]_i[\hat{\bbm[\epsilon]}_{p'}^{+}(\mathbf{k}',\omega)]_j^*\mathcal{P}_{-1}^{\text{(pw)}}\mathcal{R}^\dag\\
&\,+e^{i(k_z-k_z^{'*})z}[\hat{\bbm[\epsilon]}_p^+(\mathbf{k},\omega)]_i[\hat{\bbm[\epsilon]}_{p'}^{+}(\mathbf{k}',\omega)]_j^*\\
&\,\times\Bigl(\mathcal{T}\mathcal{P}_{-1}^{\text{(pw)}}\mathcal{T}^\dag+\mathcal{R}\mathcal{P}_{-1}^{\text{(pw)}}\mathcal{R}^\dag\Bigr)\Bigr]\ket{p',\mathbf{k}'},\\
&\alpha_\text{M}(\omega)=\frac{3\pi c}{2\omega}\!\sum_{p,p'}\sum_{i,j}\frac{[\mathbf{d}_{mn}]_i^*[\mathbf{d}_{mn}]_j}{|\mathbf{d}_{mn}|^2}\int\frac{d^2\mathbf{k}}{(2\pi)^2}\int\frac{d^2\mathbf{k}'}{(2\pi)^2}\\ &\,e^{i(\mathbf{k}-\mathbf{k}')\cdot\mathbf{r}}\bra{p,\mathbf{k}}e^{i(k_z-k_z^{'*})z}[\hat{\bbm[\epsilon]}_p^+(\mathbf{k},\omega)]_i[\hat{\bbm[\epsilon]}_{p'}^{+}(\mathbf{k}',\omega)]_j^*\Bigl(\mathcal{P}_{-1}^\text{(pw)}\\
&\,-\mathcal{R}\mathcal{P}_{-1}^\text{(pw)}\mathcal{R}^\dag+\mathcal{R}\mathcal{P}_{-1}^\text{(ew)}-\mathcal{P}_{-1}^\text{(ew)}\mathcal{R}^\dag-\mathcal{T}\mathcal{P}_{-1}^\text{(pw)}\mathcal{T}^\dag\Bigr)\ket{p',\mathbf{k}'},\end{split}\end{equation}
where $i,j\in\{x,y,z\}$. A mode of the field is identified by the polarization index $p=$TE,TM, the transverse wavevector $\mathbf{k}=(k_x,k_y)$, the frequency $\omega$ and the direction of propagation along the $z$ axis $\phi=\pm$. The three-dimensional wavevector is defined as $\mathbf{K}^\phi=(\mathbf{k},\phi k_z)$, its $z$ component being $k_z=\sqrt{\frac{\omega^2}{c^2}-\mathbf{k}^2}$. The polarization vectors $\hat{\bbm[\epsilon]}^\phi_{\mathrm{TE}}(\mathbf{k},\omega)$ and $\hat{\bbm[\epsilon]}^\phi_{\mathrm{TM}}(\mathbf{k},\omega)$ appearing in eq. \eqref{alphas} are defined as $\hat{\bbm[\epsilon]}^\phi_\TE(\mathbf{k},\omega)=\frac{1}{k}(-k_y\hat{\mathbf{x}}+k_x\hat{\mathbf{y}})$ and $\hat{\bbm[\epsilon]}^\phi_\TM(\mathbf{k},\omega)=\frac{c}{\omega}(-k\hat{\mathbf{z}}+\phi k_z\hat{\mathbf{k}})$, where $\hat{\mathbf{x}}$, $\hat{\mathbf{y}}$ and $\hat{\mathbf{z}}$ are the unit vectors along the three
axes and $\hat{\mathbf{k}}=\mathbf{k}/k$. $\mathcal{P}_{-1}^\text{(pw)}$ and $\mathcal{P}_{-1}^\text{(pw)}$ are defined as $\bra{p,\mathbf{k}}\mathcal{P}_n^\text{(pw/ew)}\ket{p',\mathbf{k}'}=k_z^n\bra{p,\mathbf{k}}\Pi^\text{(pw/ew)}\ket{p',\mathbf{k}'}$, $\Pi^\text{(pw)}$ and $\Pi^\text{(ew)}$ being the projectors on the propagative ($ck<\omega$, real $k_z$) and evanescent ($ck>\omega$, purely imaginary $k_z$) sectors respectively \cite{MesAntPRA11}.

The steady state of eq. (\ref{steady state}) gives a thermal state at temperature $T$ when, for $(nm)\in\{(32),(31)\}$, $n_{\mathrm{eff}}^{(nm)}=n(\omega_{nm},T)$. This condition is trivially verified at thermal equilibrium, when $T_\mathrm{W}=T_\mathrm{M}=T$. In this case, a peculiar cancellation of $\alpha_\mathrm{W}(\omega_{nm})$ and $\alpha_\mathrm{M}(\omega_{nm})$ occurs in eq. (\ref{n effective ij}), and populations depend only on the ratios $\hbar\omega_{nm}/k_BT$, not being affected by  the body. Remarkably, this is not the case if the system is driven out of thermal equilibrium. For $T_\text{W}\neq T_\text{M}$ the steady state (\ref{steady state}) does not necessarily coincide with a thermal state, and the decay rates can be expressed as:
\begin{eqnarray}\label{transtion rate}
\begin{pmatrix}\Gamma(\omega_{nm})\\\Gamma(-\omega_{nm})\end{pmatrix}&=&\Gamma_0 (\omega_{nm})\bigl[ \alpha_\mathrm{W} (\omega_{nm})+ \alpha_\mathrm{M} (\omega_{nm})\bigr]\nonumber \\ &\times& \begin{pmatrix}1+n_{\mathrm{eff}}^{(nm)}\\\vspace{-.3cm}\\n_{\mathrm{eff}}^{(nm)}\end{pmatrix},
\end{eqnarray}
where $\Gamma_0 (\omega_{nm})=\frac{\omega_{nm}^3|\mathbf{d}_{nm}|^2}{3\pi\epsilon_0\hbar c^3}$is the spontaneous-emission rate in vacuum. The quantities $\Gamma(\pm\omega_{nm})$ are confined between their equilibrium values at $T_\mathrm{min}$ and $T_\mathrm{max}$. Moreover, they equal the ones of thermal equilibrium at an effective temperature $T_{\mathrm{eff}}^{(nm)}$, associable to each transition, defined by $n(\omega_{nm},T_{\mathrm{eff}}^{(nm)})=n_{\mathrm{eff}}^{(nm)}$ i.e.
\begin{equation}\label{effective temperature ij}T_\text{eff}^{(nm)}=\frac{\hbar\omega_{nm}}{k_B}\Bigl[\log\Bigl(1+\frac{1}{n_\text{eff}^{(nm)}}\Bigr)\Bigr]^{-1},\end{equation}
with in general $T_{\mathrm{eff}}^{(\textrm{32})}\neq T_{\mathrm{eff}}^{(\textrm{31})}$. If the state is not a thermal one, no particular thermodynamical meaning is associated to the notion of effective temperature. Nonetheless, it is a useful and elegant mathematical tool to describe the global dynamics, its meaning being the temperature for which the decay rates at equilibrium coincide with those out of equilibrium. One can then readily interpret the global dynamics in terms of {thermal-equilibrium} physics, with the strong qualitative and quantitative difference that the two transitions feel different temperatures whose values depend on the system-body distance, the geometry of the body and the interplay of these parameters with the body optical resonances. By varying the various parameters one can control separately the two effective temperatures. One can also obtain $T_{\mathrm{eff}}^{(\textrm{32})}=T_{\mathrm{eff}}^{(\textrm{31})}$ {and in this case, even if the full system is out of thermal
equilibrium, the steady atomic state is a thermal one at a temperature always between $T_\text{W}$ and $T_\text{M}$.} This subtle mechanism allows the emergence of various interesting and counterintuitive dynamical features. For example, inversion of population ordering of the two lowest-energy states $\ket{1}$ and $\ket{2}$ may occur as soon as $n_{\mathrm{eff}}^{(\textrm{32})}<n_{\mathrm{eff}}^{(\textrm{31})}$ (i.e. as as
soon as $\omega_{32}/T_{\mathrm{eff}}^{(32)}>\omega_{31}/T_{\mathrm{eff}}^{(31)}$). This can happen if $n(\omega_{32},T_\mathrm{min})<n(\omega_{31},T_\mathrm{max})$.

While the quantities $\Gamma(\pm\omega_{nm})$, $n_{\mathrm{eff}}^{(nm)}$ and $T_{\mathrm{eff}}^{(nm)}$ associated to a given transition $(nm)$ are confined between their thermal-equilibrium values at $T_\mathrm{min}$ and $T_\mathrm{max}$, this is not the case for the steady populations {$\rho_{11}(\infty)$ and $\rho_{22}(\infty)$.} For example, it can be shown that the maximum of $ \rho_{11}(\infty)$ (minimum of $ \rho_{22}(\infty)$), obtained when $n_{\mathrm{eff}}^{(32)}=n(\omega_{32}, T_\mathrm{max}) $ and $n_{\mathrm{eff}}^{(31)}=n(\omega_{31},T_\mathrm{min}) $, is larger (smaller) than its value when $T_\mathrm{W}=T_\mathrm{M}=T_\mathrm{min}$. This can be understood since the transition $(32)$ to which the maximal temperature is associated, $T_{\mathrm{eff}}^{(\textrm{32})}=T_\mathrm{max}$, is more reactive than the transition $(31)$ to which the minimal temperature is associated, $T_{\mathrm{eff}}^{(\textrm{31})}=T_\mathrm{min}$. Equivalently, the minimum of $ \rho_{11}(\infty)$ (maximum of $ \rho_{22}(
\infty)$) is obtained when $T_{\mathrm{eff}}^{(\textrm{32})}=T_\mathrm{min}$ and $T_{\mathrm{eff}}^{(\textrm{31})}=T_\mathrm{max}$.

\section{Numerical analysis}

In the following we consider a specific example providing a numerical investigation of the peculiar effects described above for an arbitrary body \footnote{All numerical calculations refer to isotropic dipoles.}. As body we consider a slab of finite thickness $\delta$. In this case the reflection and transmission operators are diagonal in the $(\mathbf{k},\omega)$ basis. Their matrix elements are given by the Fresnel reflection and transmission coefficients modified by the finite thickness $\delta$ of the slab:
\begin{equation}\label{rhotau}\begin{split}\rho_{p}(\mathbf{k},\omega)&=r_{p}(\mathbf{k},\omega)\frac{1-e^{2ik_{zm}\delta}}{1-r_{p}^2(\mathbf{k},\omega)e^{2ik_{zm}\delta}},\\
\tau_{p}(\mathbf{k},\omega)&=\frac{t_{p}(\mathbf{k},\omega)\bar{t}_{p}(\mathbf{k},\omega)e^{i(k_{zm}-k_z)\delta}}{1-r_{p}^2(\mathbf{k},\omega)e^{2ik_{zm}\delta}}.\\\end{split}\end{equation}
Here $r_{p}(\mathbf{k},\omega)$, $t_{p}(\mathbf{k},\omega)$ and $\bar{t}_{p}(\mathbf{k},\omega)$ are the ordinary Fresnel coefficients, $k_{zm}=\sqrt{\frac{\omega^2}{c^2}\varepsilon(\omega)-\mathbf{k}^2}$ is the $z$ component of the wavevector inside the medium and $\varepsilon(\omega)$ is the dielectric permittivity of the body. For this specific case, the general formulas for $\alpha_\mathrm{W}(\omega_{nm})$ and $\alpha_\mathrm{M}(\omega_{nm})$ in terms of reflection and transmission operators reduce to
\begin{equation}\label{alphaE and alphaM}\begin{split}
\alpha_\mathrm{W}(\omega_{nm})&=\frac{\hat{1}+\mathbf{B}(\omega_{nm})+2\mathbf{C}(\omega_{nm})}{2}\cdot \mathbf{\tilde{d}}_{nm},\\
\alpha_\mathrm{M}(\omega_{nm})&=\frac{\hat{1}-\mathbf{B}(\omega_{nm})+2\mathbf{D}(\omega_{nm})}{2}\cdot \mathbf{\tilde{d}}_{nm},
\end{split}\end{equation}
where $\mathbf{\tilde{d}}_{nm}=(|[\mathbf{d}_{nm}]_x|^2,|[\mathbf{d}_{nm}]_y|^2,|[\mathbf{d}_{nm}]_z|^2)/|\mathbf{d}_{nm}|^2 $,  $\hat{1}=(1,1,1)$ and \\
\begin{equation}\label{BCD}\begin{split}
\mathbf{B}(\omega)&=\frac{3c}{4\omega}\sum_p\int_0^{\frac{\omega}{c}}\!\frac{k\,dk}{k_z}\mathbf{M}_p^+(k)\bigl(|\rho_{p}(k,\omega)|^2+|\tau_{p}(k,\omega)|^2\bigr),\\
\mathbf{C}(\omega)&=\frac{3c}{4\omega}\sum_p\int_0^{\frac{\omega}{c}}\!\frac{k\,dk}{k_z}\mathbf{M}_p^-(k)\Rea\bigl(\rho_{p}(k,\omega)e^{2ik_zz}\bigr),\\
\mathbf{D}(\omega)&=\frac{3c}{4\omega}\sum_p\int_{\frac{\omega}{c}}^{\infty}\!\!\!\frac{k\,dk}{\Ima(k_z)}e^{-2\Ima(k_z)z}\mathbf{M}_p^+(k)\Ima(\rho_{p}(k,\omega)),\\\end{split}\end{equation}
being $\mathbf{M}_1^\phi=(1,1,0)$ and $\mathbf{M}_2^\phi=\frac{c^2}{\omega^2}(\phi|k_z|^2,\phi|k_z|^2,2 k^2)$.

In eq. \eqref{BCD}, $\mathbf{B}(\omega)$ and $\mathbf{C}(\omega)$ are defined in the propagative sector while $\mathbf{D}(\omega)$ in the evanescent one. $\mathbf{C}(\omega)$ and $\mathbf{D}(\omega)$ depend on the atomic position and go to zero for large $z$ while, for small $z$, $\mathbf{D}(\omega)$ diverges so that $\alpha_\mathrm{M}(\omega)$ dominates in this region. As a consequence, all the effective temperatures tend to $T_\text{M}$ in this limit (see eqs. \eqref{n effective ij} and \eqref{effective temperature ij}) and thus the atom thermalizes at the body temperature $T_\text{M}$. $\mathbf{B}(\omega)$ is independent of $z$ and then, for $z$ large enough, it is the dominating contribution in $\alpha_\mathrm{W}(\omega)$ and  $\alpha_\mathrm{M}(\omega)$. As a result, $\alpha_\mathrm{W}(\omega)>\alpha_\mathrm{M}(\omega)$ in this region and the position of each $T_\text{eff}^{(nm)}$ in the interval $[T_\text{min},T_\text{max}]$ is governed by the value of $\mathbf{B}(\omega)$. It follows that it always
exists a distance $z$ for which $\alpha_\text{W}(\omega_{nm})=\alpha_\text{M}(\omega_{nm})$. This point delimits the two zones of influence where each temperature dominates for that specific transition. We finally observe that at thermal equilibrium $\mathbf{B}(\omega)$ does not contribute, since all the quantities are proportional to the sum $\alpha_\text{W}(\omega_{nm})+\alpha_\text{M}(\omega_{nm})$, independent of $\mathbf{B}(\omega)$ and asymptotically equal to 1.

For the numerical simulations, we consider a slab made of silicon carbide (SiC). Its dielectric permittivity is described using a Drude-Lorentz model \cite{Palik98}, implying a resonance at $\omega_r=1.495\times10^{14}\,\mathrm{rad}\,\text{s}^{-1}$ and a surface {phonon-polariton} resonance at $\omega_p=1.787\times10^{14}\,\mathrm{rad}\,\text{s}^{-1}$. A relevant length scale in this case is $ c /\omega_r \simeq2\,\mu$m while a reference temperature is $ \hbar \omega_r/k_B\simeq 1140\,$K.

\begin{figure}[h!]
\scalebox{0.59}{\includegraphics{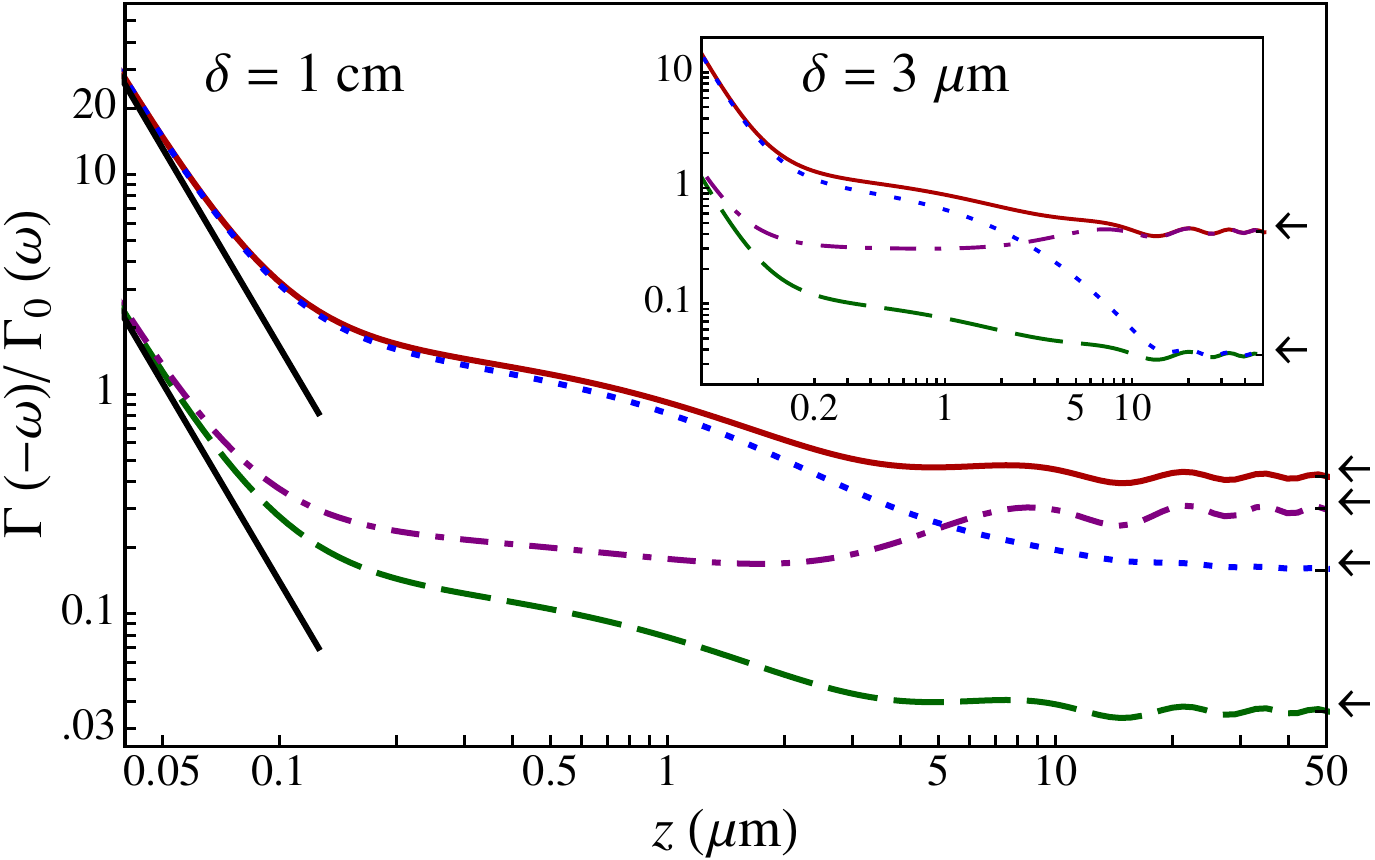}}
\caption{$\Gamma(-\omega)/\Gamma_0(\omega)$ as a function of $z$ for $\omega=0.5\omega_r$. Asymptotic curves for small $z$ (black segments) for $T_\text{W}=T_\text{M}=T_\text{max}=470\,$K and $T_\text{W}=T_\text{M}=T_\text{min}=170\,$K. $T_\text{W}=T_\text{M}=T_\text{max}$ (red solid line), $(T_\text{W},T_\text{M})=(T_\text{max},T_\text{min})$ (purple dot-dashed line), $(T_\text{W},T_\text{M})=(T_\text{min},T_\text{max})$ (blue dotted line), $T_\text{W}=T_\text{M}=T_\text{min}$ (green dashed line). The symbols $\leftarrow$ indicate the asymptotic values (with respect to $z$) corresponding to the four couples of temperatures.}
\label{Fig2}\end{figure}

In Fig. \ref{Fig2}, $\Gamma(-\omega)/\Gamma_0(\omega)$ is plotted as a function of $z$. The results out of thermal equilibrium are compared with the two cases at thermal equilibrium at $T_\text{W}$ and $T_\text{M}$. We clearly see that at small separation distances thermal-equilbrium values at $T_\text{M}$ are retrieved, while at long separation distances the influence of both $T_\text{W}$ and $T_\text{M}$ is present, as underlined by the arrows indicating {the asymptotic values}. Figure \ref{Fig2} shows that this remains true for large values of the slab thickness $\delta$. On the contrary, for values of $\delta$ comparable to the other characteristic lengths in the system, the influence of the body is limited with respect to $z$. The asymptotic curves for small $z$ diverge as $1/z^3$ \cite{BiehsPRA11}. The lack of symmetry between the case the atom approaches the slab and the case it moves far from the body results from the lack of an evanescent term in the contribution of the radiation emitted by the
walls to the local field. The evanescent contribution present in the field emitted by the slab is responsible for the fact that for $z$ small enough the slab temperature becomes dominant. For values of $z$ larger than $c/\omega \simeq 4\,\mu$m, the transition rates oscillate. This behavior is linked to the integral $\mathbf{C}(\omega)$ of eq. \eqref{BCD}, which is purely propagative, and results from the phase change of the reflected field \cite{BiehsPRA11}.

\begin{figure}[h!]
\scalebox{0.1}{\includegraphics{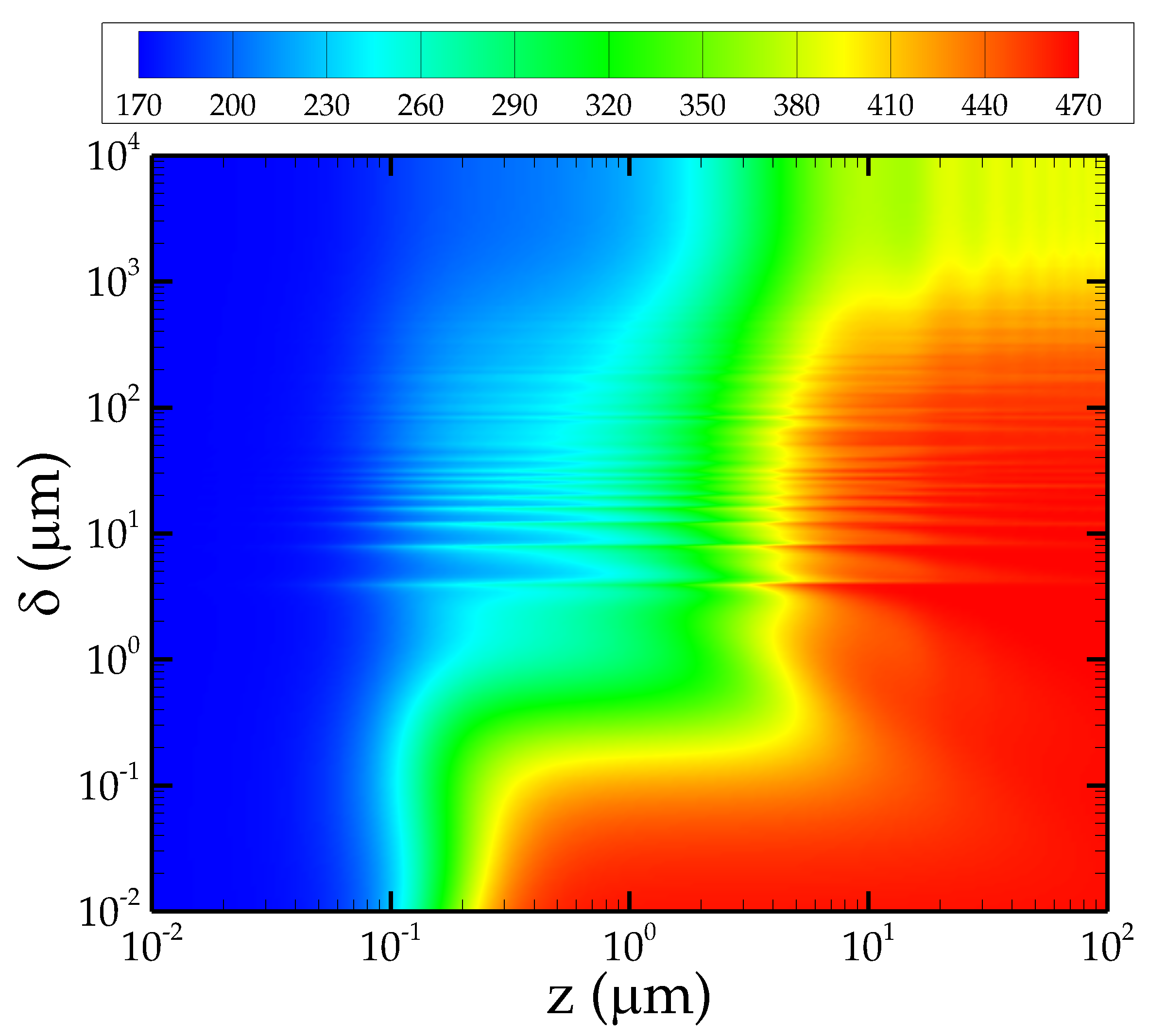}}
\caption{Density plot of $T_\text{eff}^{(nm)}$ as a function of $z$ and $\delta$. The frequency of the considered atomic transition is $\omega_{nm}=\omega_r/2$, $T_\text{M}=170$ K and $T_\text{W}=470$ K.}
\label{Fig3}\end{figure}

As a general remark, we can then state that by increasing the slab thickness $\delta$ the region of influence of the slab temperature becomes larger. The atom-slab distance determines which temperature will be more relevant in the atomic dynamics. This behavior is retrieved in Fig. \ref{Fig3}, where we plot the effective temperature $T_\text{eff}^{(nm)}$ as a function of $z$ and $\delta$ for a given frequency $\omega_{nm}$.
As previously discussed, for any thickness $\delta$, the atomic temperature for small $z$ tends always to the body temperature $T_\text{M}$, while for large distances the value of $T_\text{eff}$ results from the interplay between thickness and distance, particularly pronounced in absence of resonance between atomic transition frequency and body optical resonances. Figure \ref{Fig3} also shows the existence of regions of oscillatory behavior, originating from the propagative part of the spectrum. By varying the frequency a different dependence on $\delta$ and $z$ is obtained. One can then find couple of frequencies and values of $\delta$ e $z$ such that effective temperatures associated to the two transitions are quite different, as explicitly shown in Fig. \ref{fig:5}.

\begin{figure}[h!]
\includegraphics[width=0.49\textwidth]{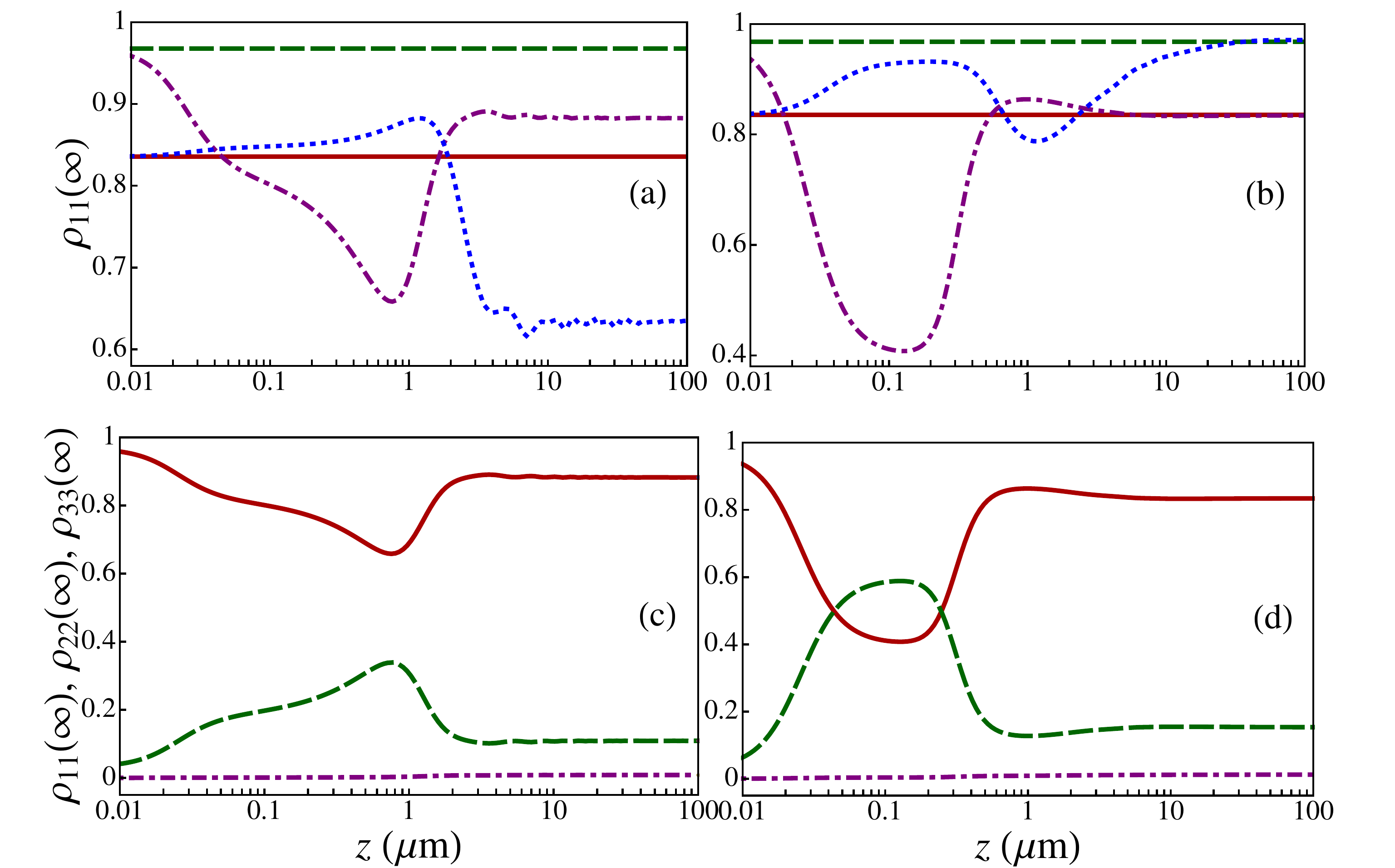}
\caption{\label{fig:4}\footnotesize {Steady populations as a function of  $z$. For all panels:} $\omega_{32}=\omega_{p} $ and $\omega_{31}=2 \omega_{r}$. Left column: $\delta = 1$ cm. Right column: $\delta = 110$ nm. Panels (a)-(b): same color convention of Fig. \ref{Fig2}  with $T_\mathrm{min}=270\,$K and $T_\mathrm{max}=540\,$K. Panels (c)-(d): $\rho_{11}(\infty)$ (red solid line), $\rho_{22}(\infty)$ (green dashed line) and $\rho_{33}(\infty)$ (dot-dashed purple line) are plotted as a function of $z$ in $\mu$m for $T_\mathrm{W}=540\,$K and $T_\mathrm{M}=270\,$K. }
\end{figure}

In Figs. \ref{fig:4}(a)-(d) we show that non-equilibrium steady populations may exceed their thermal-equilibrium values, independent of the atom-slab distance. Moreover, in the limit of semi-infinite slab ($\delta=1$ cm) the populations at {large distance} differ from the ones at thermal equilibrium at $T_\text{W}$. In this limit both $T_\mathrm{W}$ and $T_\mathrm{M}$ contribute to the transition rates. On the contrary, at small distances the populations always coincide with the ones of a thermal state at  $T_\text{M}$. The occurrence of ordering inversion of the populations $\rho_{11}(\infty)$ and $\rho_{22}(\infty)$ is depicted in panel (d). We note that the state {$\ket{3}$} is always poorly populated being $T_\mathrm{max}$ smaller than $ \hbar \omega_p/k_B\simeq 1360$ K.

\begin{figure}[h!]
\includegraphics[width=0.47\textwidth]{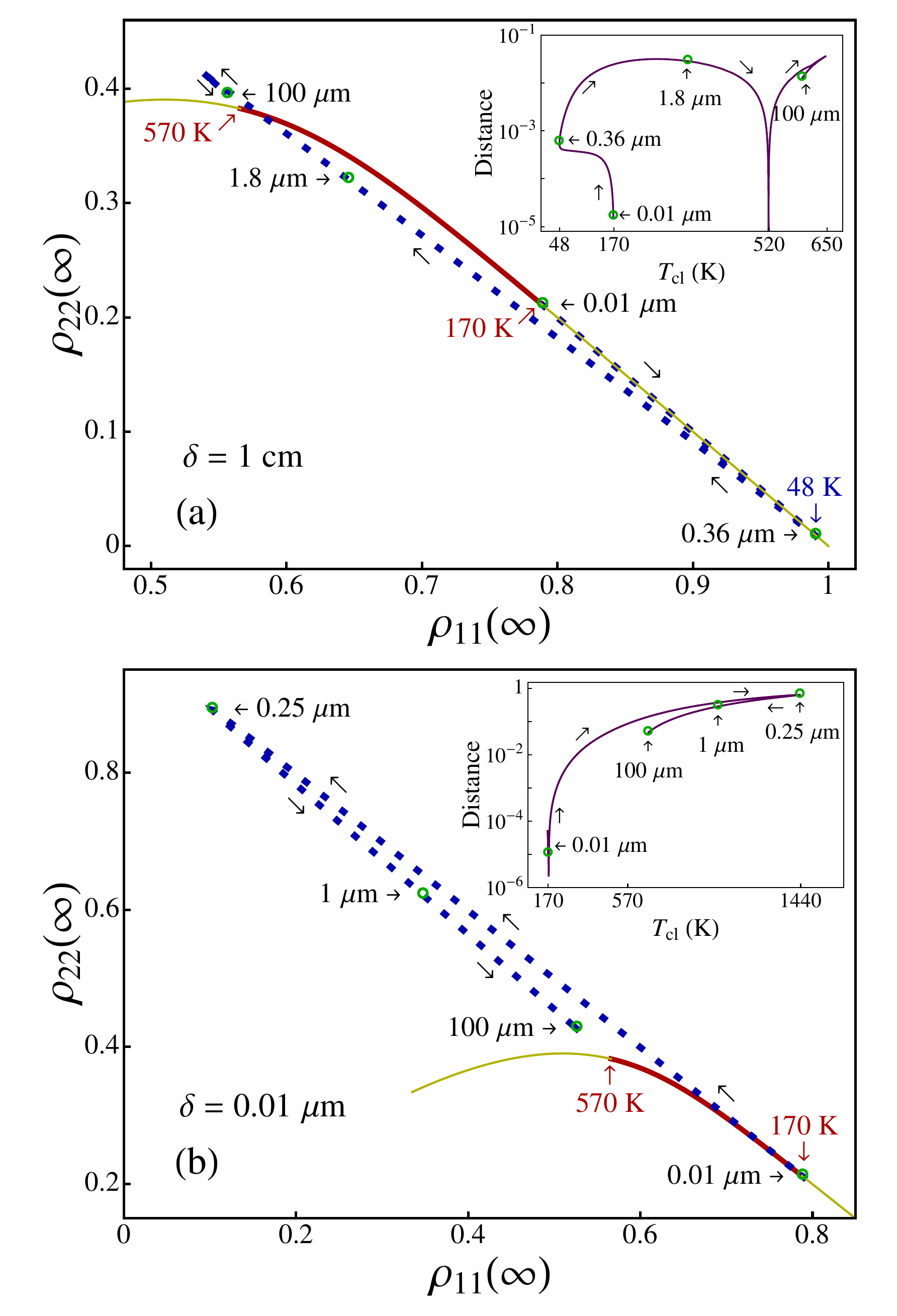}
\caption{\label{fig:5}\footnotesize For all panels: $\omega_{32}=\omega_{r} $ and $\omega_{31}=\omega_{p}$. For two values of the
slab thickness $\delta$, we plot $\rho_{22}(\infty)$ vs $\rho_{11}(\infty)$ by varying $z$ (dotted blue line) for $T_\mathrm{W}=570\,$K and $T_\mathrm{M}=170\,$K. Thermal states are represented by solid lines (170\,K$<T<570\,$K: red thick line, $T<170\,$K and $T>570\,$K: yellow thinner line). In the inset the euclidean distance from the closest thermal state is plotted as a function of its temperature $T_\text{cl}$ by varying $z$. The distances between two thermal states differing of 1\,K, respectively at $T=48\,$K, 170\,K and 570\,K, are equal to $1.3\times10^{-3}, 1.8\times 10^{-3}, 3.4\times 10^{-4}$. They may be useful to estimate how far from thermal states one is by varying $z$.}
\end{figure}

In Fig. \ref{fig:5}(a) we highlight the possibility of exploiting non-equilibrium configurations to provide an efficient atomic-cooling mechanism for its internal temperature. Indeed, for a large range of atom-slab distances $z$, the steady atomic state is practically a thermal state at a temperature {\it lower} than the minimal temperature $T_\mathrm{min}=170\,$K, as also shown by the inset reporting the distance between the atomic state and the closest thermal state,
$\sqrt{\text{Tr}(\rho-\sigma)^2}$. In particular, for $z=0.36\,\mu$m the atom is cooled down to an effective temperature around 48\,K (temperature of the closest thermal state).
It is remarkable that the above effect can be alternatively produced by starting from a thermal-equilibrium configuration with  $T_\mathrm{W}=T_\mathrm{M}=170\,$K, by properly fixing the atomic distance and by increasing $T_\mathrm{W}$ up to 570\,K. As a result, the effective atomic internal temperature cools down towards very low-temperature stationary states. The cooling mechanism derives from the effective temperatures, that at $z=0.36\,\mu$m are $T_{\mathrm{eff}}^{(\textrm{32})}\approx390\,$K $\gg$ $T_{\mathrm{eff}}^{(\textrm{31})}\approx178\,$K. We remark that the cooling discussed here has nothing to do with a reduction of the mean kinetic energy of the atom. It refers to the internal atomic temperature associated to the Boltzmann factors in the populations. From the inset one sees that at a certain position the steady state is exactly a thermal state (the distance from the closest thermal state goes to zero). This happens at $z\approx2.25\,\mu$m where $T_{\mathrm{eff}}^{(\textrm{32})}=T_{\mathrm{eff}}^{(\textrm{31})}\approx520\,$K. {Steady states far from thermal ones can also be achieved.} This is particularly evident in {Fig. \ref{fig:5}(b)} where we show that at $z=0.25\,\mu$m the steady state is at its largest distance from the curve of thermal states with an associated value of $\rho_{22}(\infty)$ of almost 0.9. At this position the effective temperatures are inverted, with $T_{\mathrm{eff}}^{(\textrm{32})}\approx227\,$K and $T_{\mathrm{eff}}^{(\textrm{31})}\approx476\,$K. Figure \ref{fig:5} elucidates how non-equilibrium configurations provide new tools to realize a large variety of different steady states.

\section{Conclusions}

In conclusion, we have investigated the thermalization mechanism of a three-level atomic quantum system placed close to an arbitrary body whose temperature is different from that of the walls surrounding the atom-body system. We provide closed-form expressions for atomic decay rates in terms of the scattering matrices of the body valid for {\it arbitrary geometrical and material properties} in systems both in and out of thermal equilibrium. We show that configurations out of thermal equilibrium can be exploited to obtain a large variety of steady states, both thermal and non-thermal, with populations that can significantly differ from their corresponding values at thermal equilibrium. Differently from the case of thermal equilibrium, they depend on all the parameters characterizing the atomic system and its environment. In the case the body is a slab, its thickness regulates the extension of the zones of influence of the two involved temperatures, while the atomic position determines which temperature is
more relevant in the atomic dynamics. Thermalization dynamics can be interpreted in terms of effective temperatures associated to each transition. We predict peculiar behaviors such as ordering inversion of the populations and cooling of  the effective atomic internal temperature, based on steady configurations without any additional external laser source. Our predictions can be relevant for a wide class of experimental configurations involving different physical realizations of elementary quantum systems, as real or artificial atoms such as quantum dots. A possible experimental realization should contain an efficient mechanism to keep the ``atom" at a given average location. For example, one can imagine to employ a trapped BEC \cite{ObrechtPRL07}, or an atomic beam \cite{HindsPRL93} for ultracold gases, and a mechanical suspension \cite{Saidi09} for quantum dots.

\acknowledgements {B. B. and M. A. acknowledge G. Compagno for useful discussions, and acknowledge financial support from the Julian Schwinger Foundation.}

\end{document}